\documentclass{appolb}
\usepackage{epsfig}

\begin{document}
\title{BOSE-EINSTEIN CORRELATIONS AS REFLECTION OF CORRELATIONS OF
FLUCTUATIONS}
\author{O.V.Utyuzh, G.Wilk
\address{The Andrzej So\l tan Institute of Nuclear Studies, Warsaw, Poland}
\and  Z. W\l odarczyk
\address{Institute of Physics, \'Swi\c{e}tokrzyska Academy, Kielce, Poland}
}
\maketitle
\centerline{\it Dedicated to Stefan Pokorski in honour of his 60th birthday}
\vspace{2mm}

\begin{abstract}
Bose-Einstein correlations (BEC) observed between identical bosons
produced in high energy multiparticle collisions are regarded as very
important tool in investigations of multiparticle production
processes. We present here their stochastic feature stressing the
fact that they can be regarded as a reflection of correlations of
fluctuations present in hadronizing system. We show in particular
that such approach allows for simple modelling of BEC in numerical
event generators used to describe the multiparticle production
processes at high energy collisions. 

\end{abstract}

\section{Introduction}
Bose-Einstein correlations (BEC) between identical bosons are since
long time recognized as very important tool in searching for dynamics
of multiparticle production processes because of their ability to
provide the space-time information about them \cite{BEC}. This is
particularly important for heavy ion collisions which are expected to
provide us with the new state of matter, the Quark Gluon Plasma (QGP)
\cite{QGP}. However, because of their complexity, all these processes
can be investigated only by numerical modelling methods using
different sorts of Monte Carlo (MC) event generators \cite{GEN}.
Their {\it a priori} probabilistic structure prevents occurring of
genuine BEC which are of purely quantum statistical origin. The best
one can do is to {\it model} BEC by changing the outputs of these
generators in such a way as to reproduce the characteristic signals
of BEC obtained experimentally. In the most widely investigated case
of $2$-particle BEC it is the fact that two-particle correlation
function 
\begin{equation}
C_2(Q=\vert p_i - p_j\vert )\, =\,
             \frac{N_2(p_i,p_j)}{N_1(p_i)\, N_1(p_j)} . \label{eq:C2}
\end{equation}
defined as ratio of the two-particle distributions to the product of
single-particle distributions increases towards $C_2 = 2$ when $Q$
approaches zero.

\section{More about BEC}
\subsection{BEC - space-time approach}
There are two possible approaches towards BEC. The first stresses
their space-time features and is based on the symmetrization of the
respective multiparticle wave function \cite{BEC} expressed by plain
waves\footnote{This is idealization neglecting both the possible
final state and Coulomb interactions inclusion of which is possible
by a suitable modifications of these plane waves. We shall not
discuss it here.}, $ e^{\pm ikx}$, representing the produced
particles. After 
symmetrization (and squaring) one gets the respective many-particle
production rates depending on combination of variables of the type:
$(k_i-k_j)(x_i - x_j)$. To get $C_2$ as given by eq.(\ref{eq:C2}),
one has to integrate them, with some assumed weight function
$\rho(x_1,x_2,\dots)$, over unmeasured space-time positions $\{x_i\}$
of the production points. The distribution $\rho(x_1,x_2,\dots,)$ is
customarily assumed to be separable in terms of single particle
distributions $\rho_i(x_i) = \rho(x)$ and in this way the information
on the space-time distribution of points of production of finally
observed particles enters here. It can be then show that, under some
assumptions \cite{BEC}, 
\begin{equation}
C_2(Q) = 1 + \left | \int dx\rho(x)\cdot e^{iQx} \right | ^2 = 1 + \mid
\tilde{\rho}(Q)\mid ^2 , \label{eq:ftransform}
\end{equation}
i.e., $C_2(Q)$ can be regarded as a (kind of) Fourier transform of
the space-time dimensions of the emitting source\footnote{Actually,
after closer inspection \cite{ZAJC} it turns out that one rather
gets in this way a Fourier transform of the distributions of
two-particle separations (or {\it correlation lengths} \cite{BEC}).}.
So far this approach is dominating in what concerns description of
BEC.  

\subsection{BEC - quantum-statistical approach}
The second approach is based on observation that one encounters
similar correlations in quantum optics \cite{OPTICS} where they are
known as the so called HBT effect. They are described there as
arising because of correlations of some specific fluctuations present
in physical systems considered (known as {\it photon bunching} effect
\cite{OPTICS}). Following \cite{ZAJC,F} one can apply such
possibility to description of hadronizing sources as well. Because  
\begin{equation}
\langle n_1 n_2\rangle =  \langle n_1\rangle \langle n_2\rangle
                     + \langle \left(n_1 - \langle n_1\rangle\right)
               \left(n_2 - \langle n_2\rangle\right)\rangle =
                \langle n_1\rangle \langle n_2\rangle
                 + \rho \sigma(n_1)\sigma(n_2) \label{eq:COV}
\end{equation}
(where $\sigma(n)$ is dispersion of the multiplicity distribution
$P(n)$ and $\rho$ is the correlation coefficient depending on the
type of particles produced: $\rho = +1,-1,0$ for bosons, fermions and
Boltzmann statistics, respectively) one can write two-particle
correlation function (\ref{eq:C2}) in terms of the above covariances
(\ref{eq:COV}) stressing therefore its stochastic character:
\begin{equation}
C_2(Q=|p_i-p_j|) = \frac{\langle
n_i\left(p_i\right)n_j\left(p_j\right)\rangle}
  {\langle n_i\left(p_i\right)\rangle\langle
n_j\left(p_j\right)\rangle}
       = 1 + \rho \frac{\sigma\left(n_i\right)}
                             {\langle n_i\left(p_i\right)\rangle}
                        \frac{\sigma\left(n_j\right)}
                        {\langle n_j\left(p_j\right)\rangle} .
\label{eq:algor}
\end{equation}
It means therefore that $C_2(Q)$ can be regarded as being a measure
of correlation of fluctuations. This fact has been used for numerical
modelling of BEC in \cite{OMT} where a special MC generator, based on
application of information theory, was constructed for this purpose.
In it the identical pions produced in a given event were bunched on a
maximal possible way (restricted only by conservation laws
constraints) in a limited number of {\it elementary emitting cells}
of phase space according to Bose-Einstein distribution, $P(E_i) \sim
\exp \left[ n_i \left(\mu - E_i\right)/T \right]$ ($n_i$ is their
multiplicity and $E_i$ are their energies)\footnote{Values of two
lagrange multipliers, $T$ and $\mu$, were fixed by the
energy-momentum and charge conservation constraints, respectively.
Such distribution represents typical example of nonstatistical
fluctuations present in the hadronizing source. Similar concept of
elementary emitting cells has been also proposed in \cite{BSWW}.},
with size (in rapidity, as only one dimensional phase space was
considered) given by parameter $\delta y$. It turns out that in this
approach one gets at the same time both the correct BEC pattern
(i.e., correlations) and fluctuations (as characterized by the
observed intermittency pattern) \cite{OMT}. This is very strong
advantage of this model, which is so far the only example of
hadronization model, in which Bose-Einstein statistics is not only
included from the very beginning on a single event level, but it is
also properly used in getting the final secondaries. In all other
approaches \cite{LS}-\cite{AFTER} at least one of the above elements
is missing. The shortcoming of method \cite{OMT} are numerical
difficulties to keep the energy-momentum conservation as exact as
possible and its limitation to the specific event generator only.   

\subsection{Existing methods of numerical modelling of BEC}
In all other approaches the effect of BEC is obtained by a suitable 
changing the original output of MC generators used and introducing
this way (more or less artificially) desired bunching in the
phase-space of the finally produced identical particles \cite{LS,FW} 
\footnote{The specific approaches proposed for LUND model  \cite{AR}
and the afterburner method discussed in \cite{AFTER}, which we shall
not discussed here, also belong here.}. This is achieved either by
$(a)$ shifting (in each event) momenta of adjacent like-charged
particles in such a way as to get desired $C_2(Q)$ \cite{LS} (one has
to correct afterwards for the energy-momentum imbalance introduced
this way), or by $(b)$ screening all events obtained from a particular
MC generator against the possible amount of bunching they are already
showing and counting them as many times as necessary to get
desirable $C_2(Q)$ \cite{FW} \footnote{Technically this is realised 
by multiplying each event by a special weight calculated using the
output provided by event generator used.}. The original
energy-momentum balance remains in this case intact whereas the
original single particle distributions are changed (this fact can be
corrected by running again generator with suitably modified input
parameters). In both cases one uses specific weights constructed from
the assumed shape of $\rho(x)$ functions. However, the size
parameters occuring there bear no direct resemblance to the size
parameter $R$ obtained by directly fitting data on $C_2(Q)$ in
eq.(\ref{eq:C2}) using simple gaussian or exponential forms. They
rather represent instead the corresponding correlation lengths
between the like particles \cite{BEC}. 

\section{Numerical modelling of BEC understood as correlations of
fluctuations} 
Recently we have proposed \cite {NEWBEC} a new method of numerical
modelling of BEC understood as manifestation of correlations of
fluctuations, which applies already on a single event level, does not
violate any conservation laws and can be applied to data provided by
essentially any event generator modelling multiparticle production.
Here we would like to present physical ideas underlying our approach
in more detail. 

\vspace{-7mm}
\begin{figure}[h]
\noindent
\begin{center}
\epsfig{file=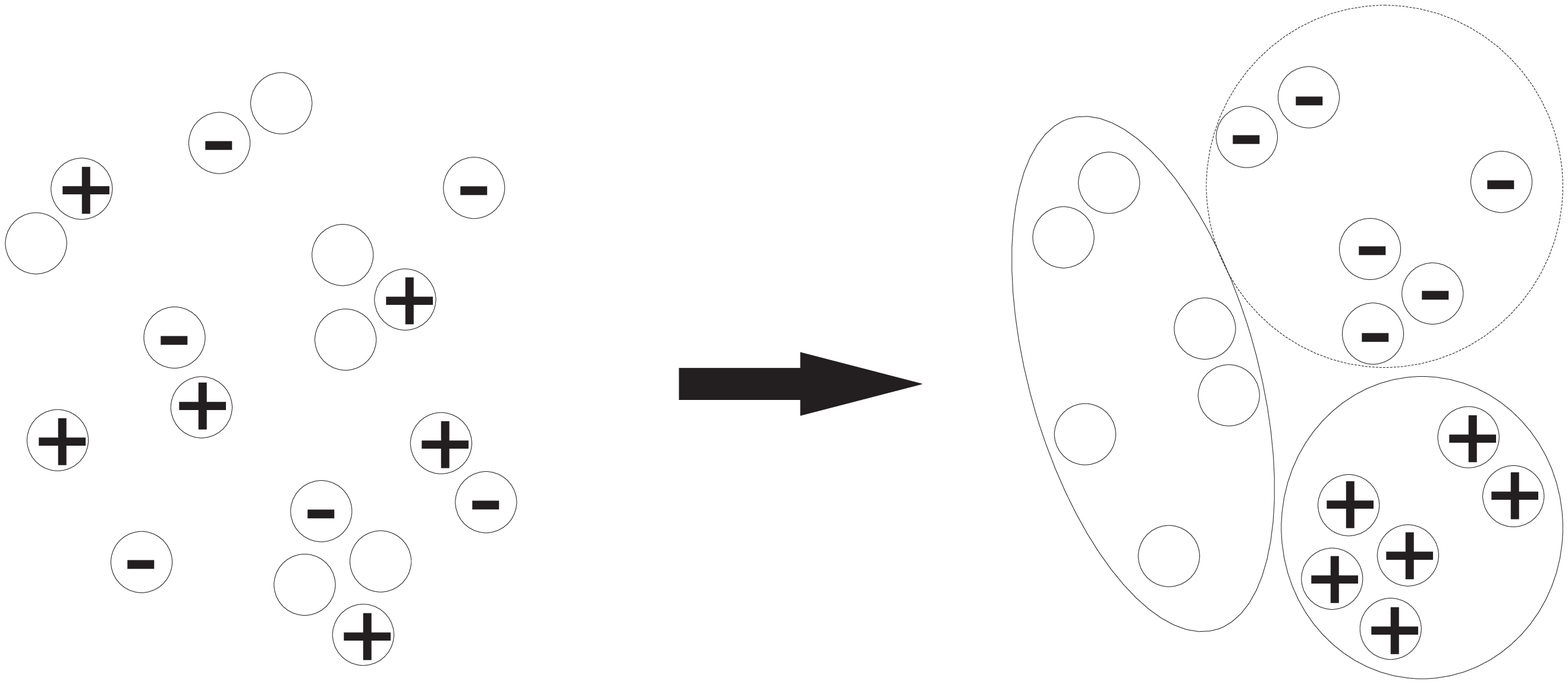, width=80mm}
\end{center}
\vspace{-10mm}
\caption{}
\label{Fig1}
\end{figure}
\vspace{-3mm}

Let us start with very simple example of what we are aiming at.
Suppose that our MC event generator provides us with a number $N(+)$,
$N(-)$ and $N(0)$ of positively and negatively charged particles 
and neutral ones located in phase space, cf. Fig. 1, left panel. They
are all uniformly distributed and show no BEC pattern. Suppose now
that {\it the same particles} (i.e., located at the same space-time
points and possessing the same momenta as before, with the same
$N(+)$, $N(-)$ and $N(0)$) have now {\it different allocation of
charges}, namely the one shown in the right panel of Fig. 1. The like
charges are in visible (albeit strongly exaggerated) way bunched
(correlated) together leading to signal of BEC. What we have done in
this example is the following: $(a)$ we have resigned from the (not
directly measurable) part of the information provided by event
generator concerning the charge allocation to produced particles  and
$(b)$ we have allocated charges anew in such a way as to keep the
like charges as near in phase space as possible (keeping also the
total charge of any kind the same as the original one). It is
interesting to note that this can be regarded as introduction of
quantum mechanical element of uncertainty to the otherwise classical
scheme of MC generator used (however, it differs completely from the
usual attempts to introduce quantum mechanical effects discussed in
\cite{QUANT}). 

That such simple scheme really works can be seen in Fig. 2, which
shows the $C_2(Q)$ for one dimensional lattice of $N$ pions
(positive, negative and neutral) with momenta $p_i = -p_{max} + 
(i - 1)\cdot \Delta p$ where spacing $\Delta p = 2p_{max}/N)$. When
their charges are assigned in a purely random way (what 
\begin{figure}[ht]
\noindent
  \begin{minipage}[ht]{38mm}
    \centerline{
        \epsfig{file=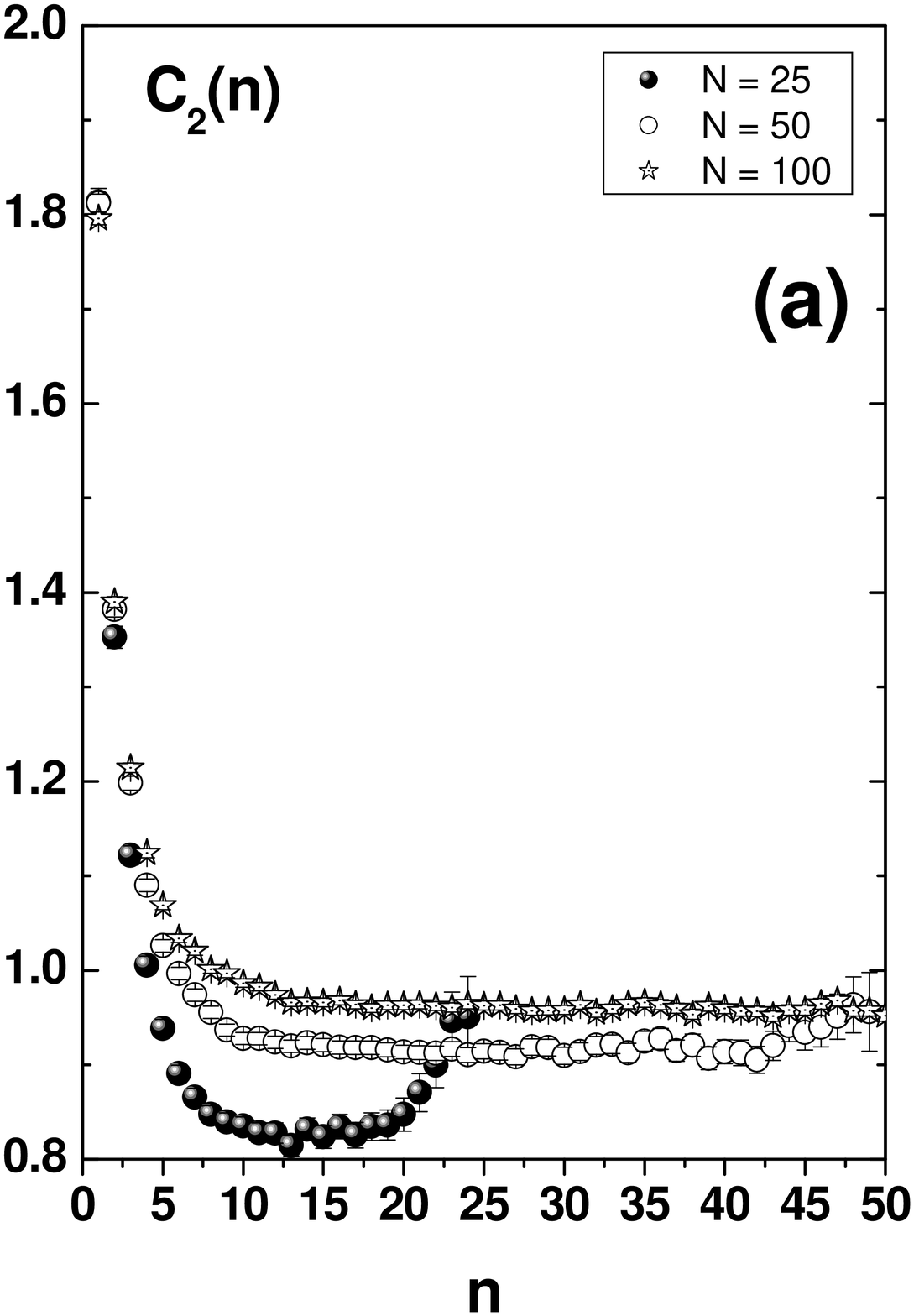, width=38mm}
     }
  \end{minipage}
\hfill
  \begin{minipage}[ht]{38mm}
    \centerline{
       \epsfig{file=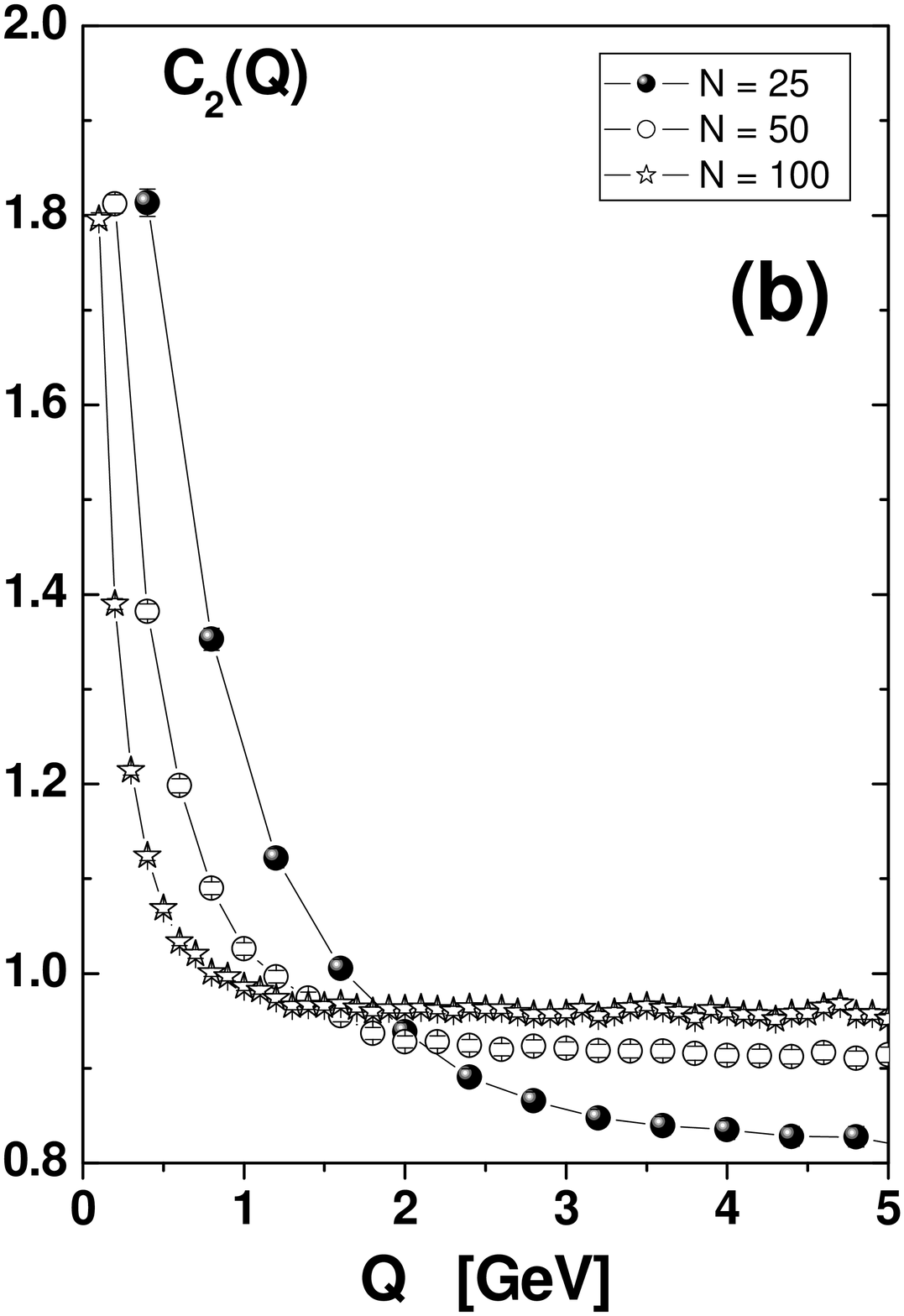, width=38mm}
     }
  \end{minipage}
\hfill
  \begin{minipage}[ht]{38mm}
    \centerline{
       \epsfig{file=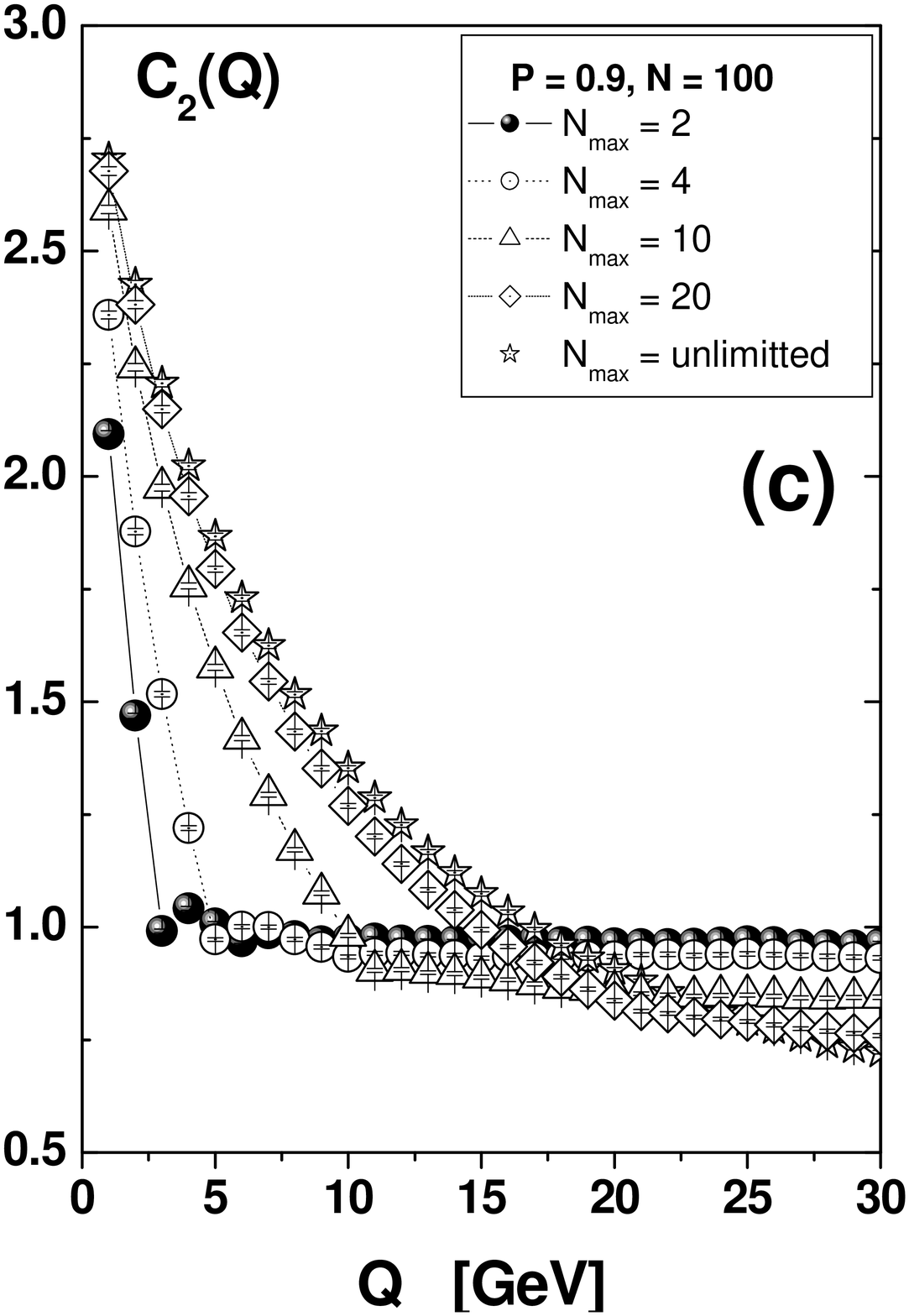, width=38mm}
     }
  \end{minipage}
  \caption{Example of $C_2$ occuring for a pionic lattice in
(one-dimensional) momentum space, $Q=|p_i - p_j|= 2p_{max}|i-j|/N =
2p_m n/N$ : $(a)$ as a function of $n$; $(b)$ as a function of $Q$
; $(c)$ as a function of $Q$ but for limitation of cell occupancy to
$i<N_{max}$. In all cases $p_{max} = 10$ GeV). In $(a)$ and $(b)$
$P=0.5$ whereas in $(c)$ $P=0.9$ (to allow for large cells).} 
  \label{Fig2}
\end{figure}
corresponds to the situation shown at the left part of Fig. 1) it can
be shown that the corresponding $C_2(Q) =1$. However, assigning
charges in a specific way (following prescription used in
\cite{NEWBEC}) one gets strong enhancement of $C_2(Q)$ which normally
is attributed to BEC. The procedure used is very simple. First one of
the particles (from $N_{\pi}$) is selected and some charge (out of
$(+,-,0)$) is randomly allocated to it. After that the same charge is
allocated to as many particles located nearby in phase space as
possible in some prescribe way forming a cell in phase-space occupied
by particles of the same charge only (cf. right part of Fig. 1). This
process is then repeated until all particles are used. The important
point is to ensure that the above selection is done in such way as to
get geometrical (Bose-Einstein) distribution of particles in a given
cell. This can be achieved by selecting each next particle with some
fixed probability $P$ till the first failure, after which the new
cell is formed. In this case $\sigma = <n> = P/(1-P)$ and second term
in the eq.(\ref{eq:algor}) is now maximal. We refer to \cite{NEWBEC}
for details of the algorithm used. The characteristic pattern
emerging here is that the so called "radius parameter" $R$ (in the
usual fitting formula for $C_2(Q) = \gamma [1 + \exp(-R\cdot Q)]$)
increases with number of particles allocated to our lattice (i.e.,
with decreasing of their momentum separation $\Delta p$, cf. Fig. 2b
(in terms of the number of particles considered it is the same, see
Fig. 2a). On the other hand it decreases with the number of particles
one can allocate to a given cell (cf. Fig. 2c).

We shall illustrate now action of our algorithm on simple cascade
model of hadronization (CAS) (in its one-dimensional versions and
assuming, for simplicity, that only direct pions are produced)
\cite{CAS} and on equally simple model based on application of
\begin{figure}[ht]
\noindent
  \begin{minipage}[ht]{38mm}
    \centerline{
        \epsfig{file=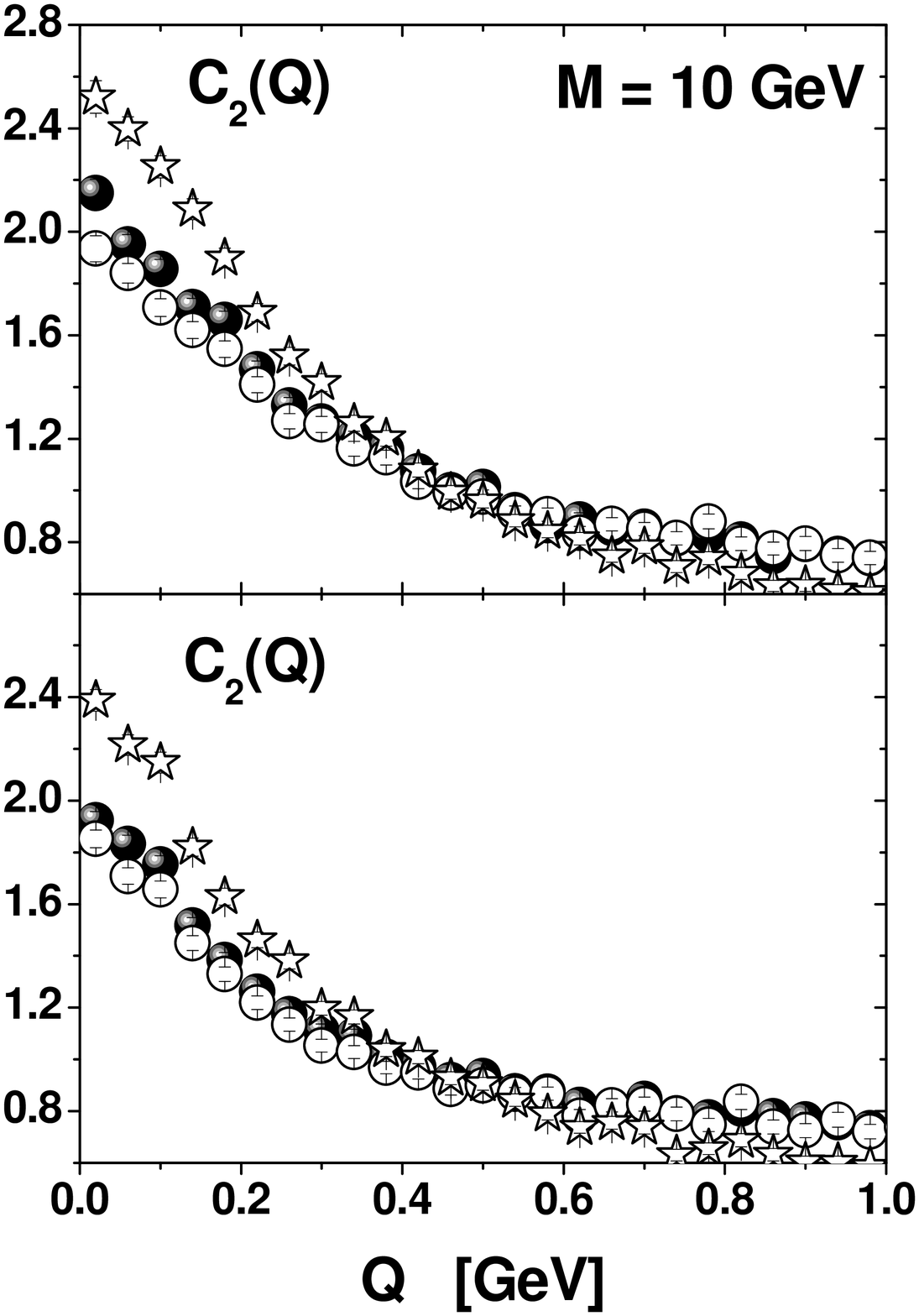, width=38mm}
     }
  \end{minipage}
\hfill
  \begin{minipage}[ht]{38mm}
    \centerline{
       \epsfig{file=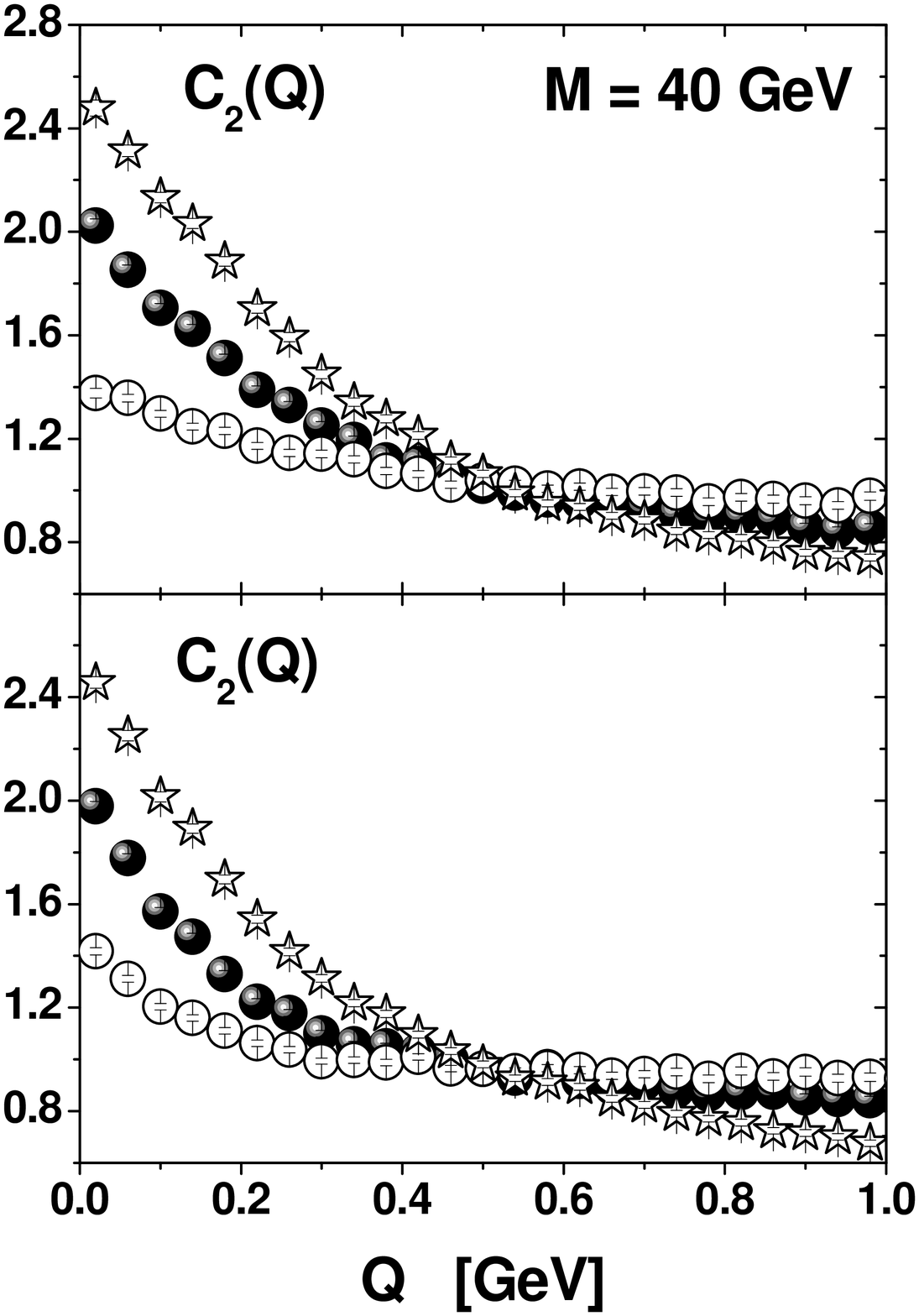, width=38mm}
     }
  \end{minipage}
\hfill
  \begin{minipage}[ht]{38mm}
    \centerline{
       \epsfig{file=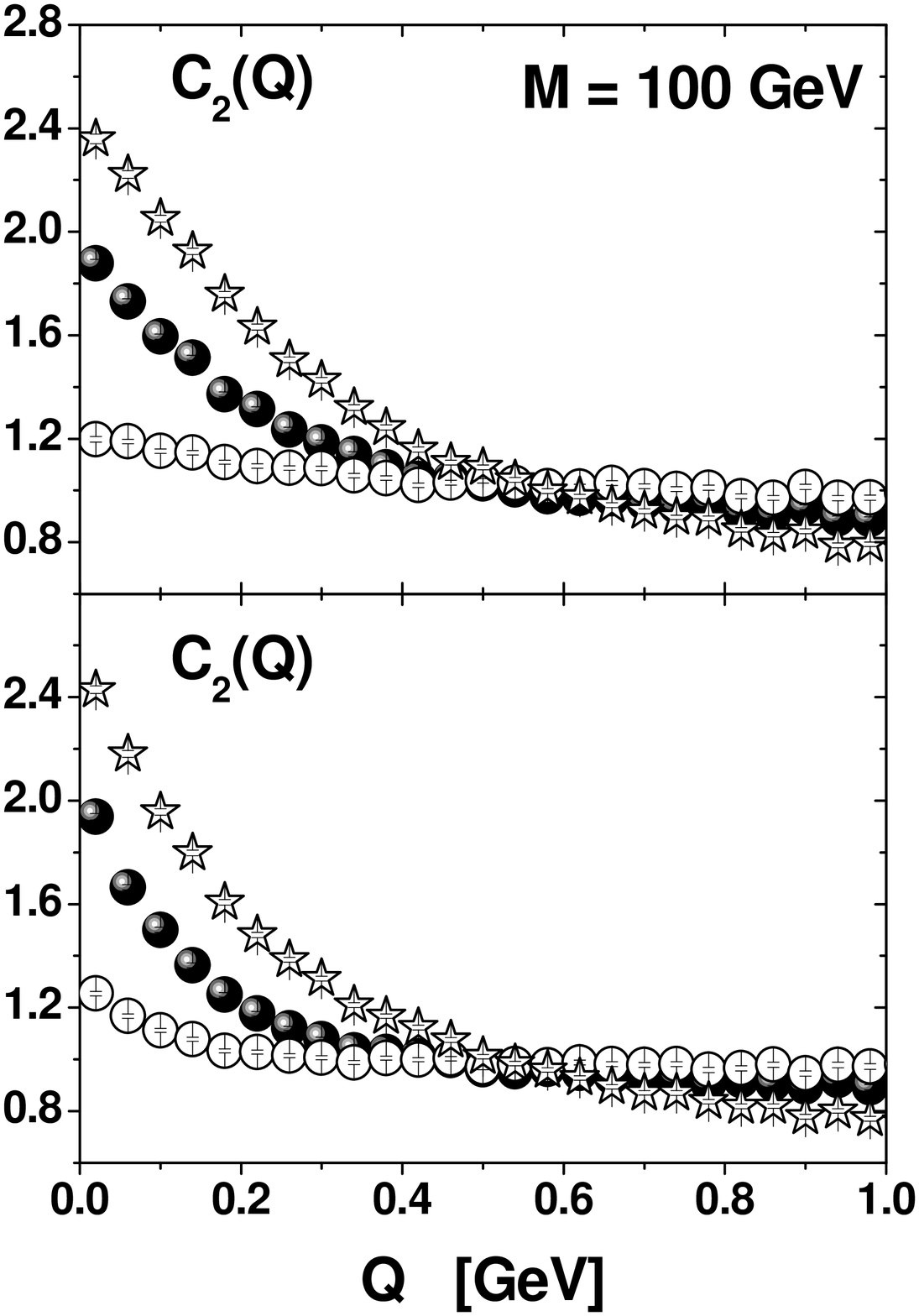, width=38mm}
     }
  \end{minipage}
  \caption{Examples of BEC patterns obtained for $M=10$,
$40$ and $100$ GeV for constant weights $P=0.75$ (stars) and $P=0.5$ 
(full symbols) and for the weight given by eq.(\ref{eq:CASMaxEnt}) 
(open symbols). Upper panels are for CAS, lower for MaxEnt (see text
for details).}
  \label{Fig3}
\end{figure}
information theory \cite{MaxEnt} (cf. Fig. 3). In CAS the initial
mass $M$  
hadronizes by series of well defined (albeit random) branchings
($M\rightarrow M_1+M_2$, with $M_{1,2} = r_{1,2}M$ such that $r_1 +
r_2 < 1$) and is endowed with a simple spatio-temporal
pattern. It shows no traces of Bose-Einstein statistics whatsoever.
In MaxEnt particles occur instanteneously in all phase space with
distribution given by the thermal-like formula obtained by
maximalization of the accordingly defined information entropy. In
both cases the masses and multiplicities were kept the same. There is
no BEC here either. However, as can be seen in Fig. 3, when endowed
with charge selection provided by our algorithm, a clear BEC pattern
emerges in $C_2(Q)$ (and is very similar in both cases considered
here). Two kind of  choices of probabilities are shown in Fig. 3.
First is constant $P=0.75$ and $P=0.5$. The other is what we
call the "minimal" weight constructed from the output information
provided by CAS ($P_M$) or MaxEnt ($P_{ME}$) event generators: 
\begin{equation}
P_{C}(ij) = \exp\left[- \frac{1}{2} \delta^2_{ij}(x)\cdot\delta^2_{ij}(p)
\right]\quad {\rm or}\quad
P_{ME}(ij) = \exp\left[ - \frac{\delta^2_{ij}(p)}{2\mu_i T_l}\right] ,
\label{eq:CASMaxEnt}
\end{equation}
where $\delta_{ij}(x) = |x_i-x_j|$, $\delta_{ij}(p) = |p_i - p_j|$
and $T_l$ is the corresponding "temperature" (with $\mu$ being mass
of the produced particles). In this way one connects $P$ with details
of hadronization  process by introducing to it a kind of overlap
between particles as a measure of probability of their bunching in a
given emitting cell. 

It turns out that BEC effect shown in Fig. 3 depends only on the
(mean) number of particles of the same charge in phase-space cell 
and on the (mean) numbers of such cells. This depends on $P$, the
bigger $P$ the more particles and bigger $C_2(Q=0)$; smaller $P$
leads to the increasing number of cells, which, in turn, results in
decreasing $C_2(Q=0)$, as already noticed in \cite{BSWW}. For small
energies the number of cells decreases in natural way while their
occupation remains the same (because $P$ is the same), therefore the
corresponding $C_2(0)$ is bigger, as seen in Fig. 3. The fact that 
there is tendency to have $C_2(0) > 2$ for larger $P$ means that one
has in this case more cells with more than $2$ particles allocated to
them, i.e., it is caused by the influence of higher order BEC.
Therefore the "sizes" $R$ obtained from the exponential fits to
results in Fig. 3 (like $C_2(Q) \sim 1 + \lambda\cdot \exp(-Q\cdot
R)$ where $\lambda$ being usually called chaoticity parameter
\cite{BEC}) correspond to the sizes of the respective elementary
cells rather than to sizes of the whole hadronizing sources itself. 
For $P=0.5$ the "size" $R$ varies weakly between $0.66$ to $0.87$ fm
from $M=10$ to $100$ GeV whereas for the "minimal" weight
(\ref{eq:CASMaxEnt})  it varies from $0.64$ to $0.44$ fm.

So far we were considering only single sources. Suppose now that
source of mass $M$ consists of a number ($n_l=2^k$) of subsources
hadronizing independently. It turns out that the resulting $C_2$'s
are very sensitive to whether in this case one applies our algorithm of
assigning charges to all particles from subsources taken together
("Split" type of sources) or to each of the subsource independently
("Indep" type of sources), cf. Fig. 4. Whereas the later case (in
which particles remember from which source they have originated)
\vspace{-3mm}
\begin{figure}[ht]
\noindent
  \begin{minipage}[ht]{38mm}
    \centerline{
        \epsfig{file=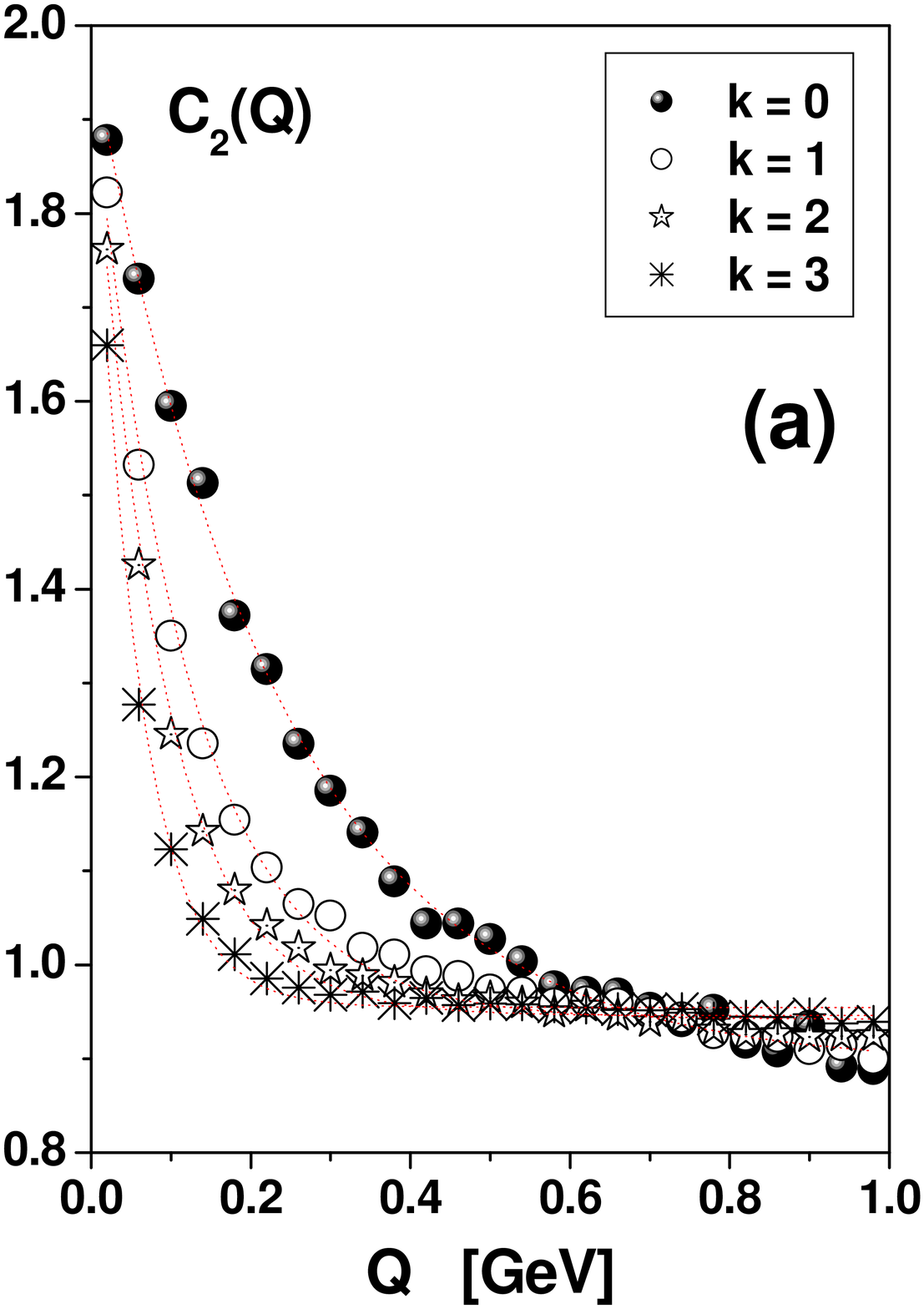, width=33mm}
     }
  \end{minipage}
\hfill
  \begin{minipage}[ht]{38mm}
    \centerline{
       \epsfig{file=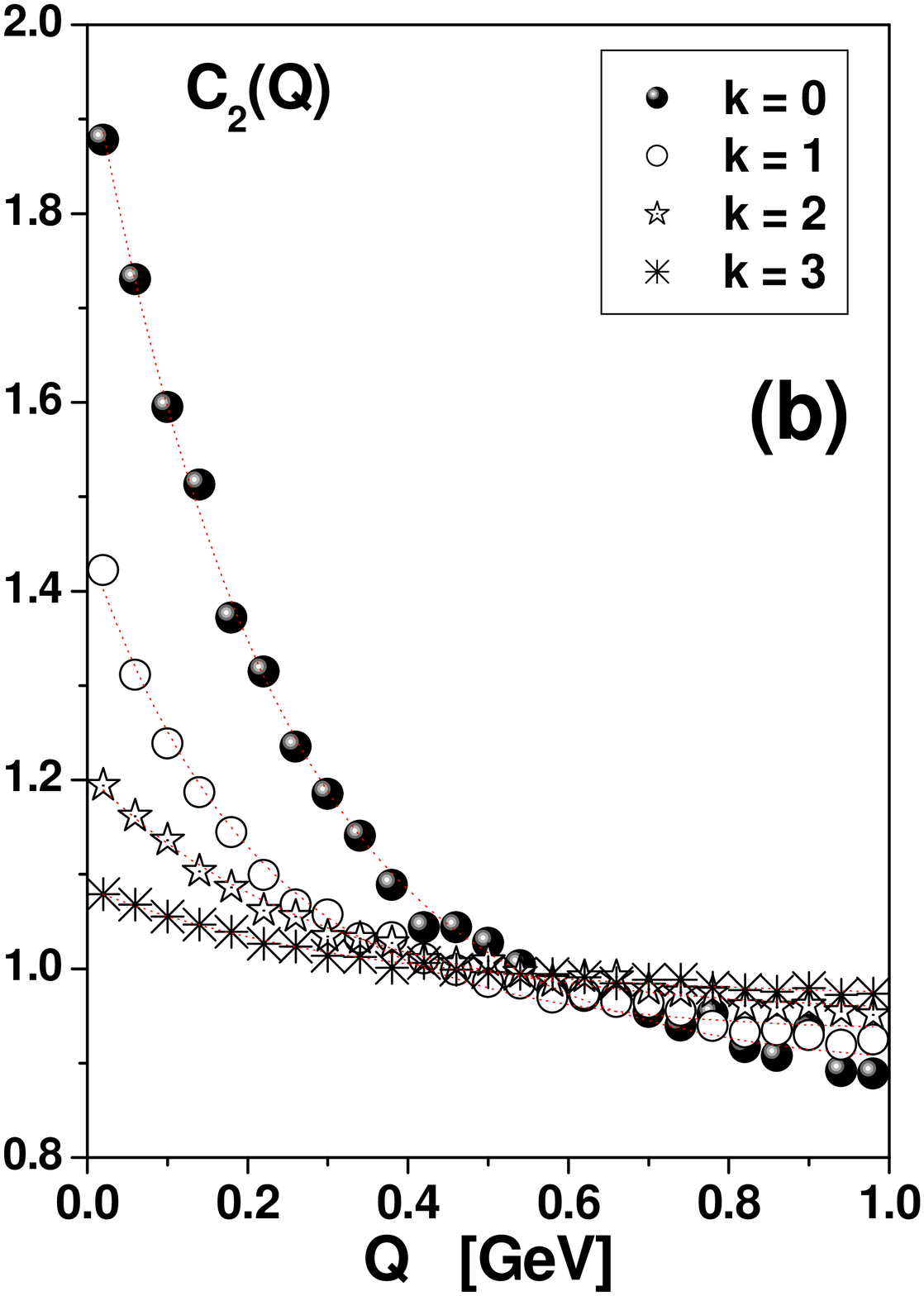, width=33mm}
     }
  \end{minipage}
\hfill
  \begin{minipage}[ht]{38mm}
    \centerline{
       \epsfig{file=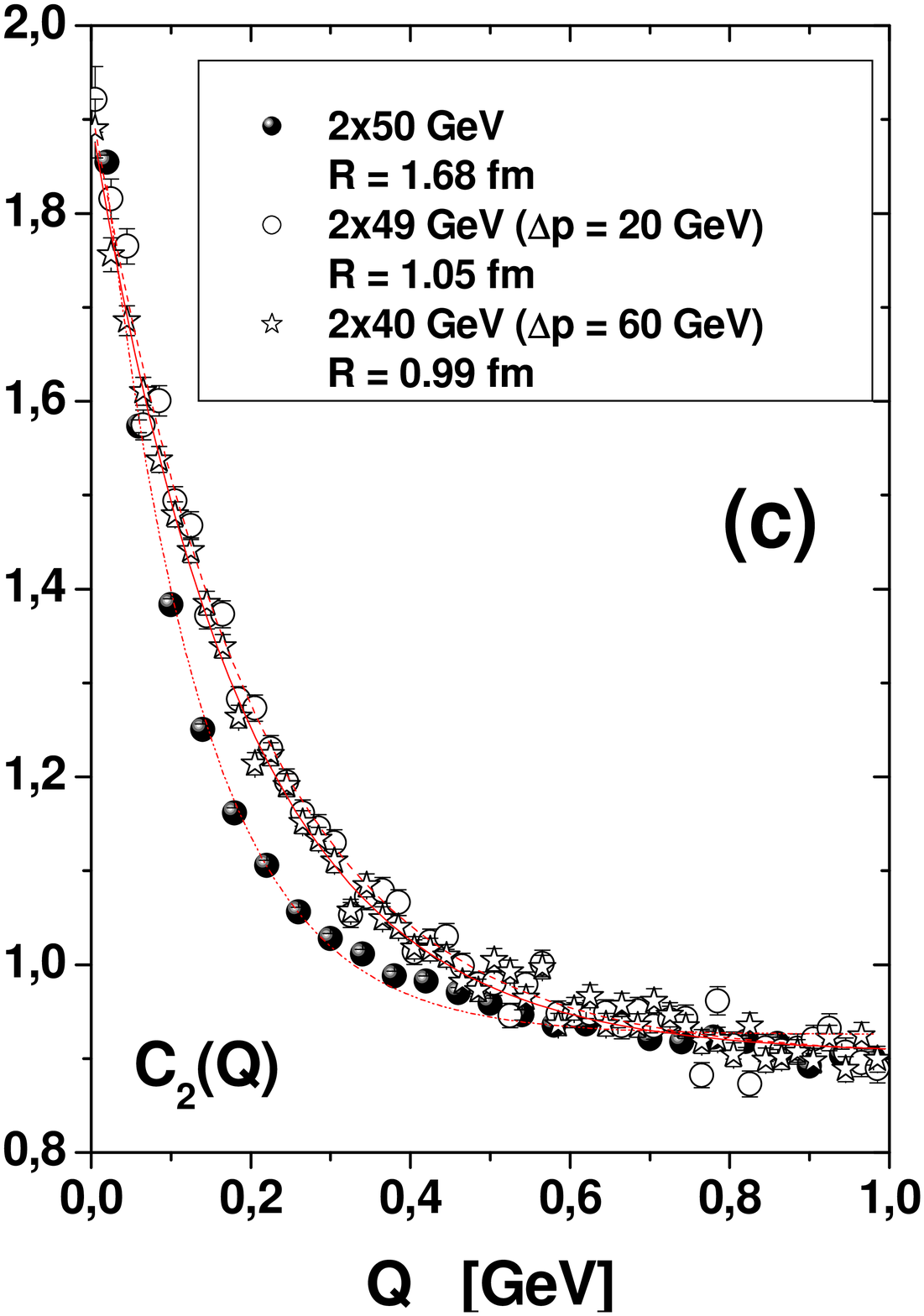, width=33mm}
     }
  \end{minipage}
  \caption{Examples of BEC for CAS model with $P=0.5$ calculated
for different number of subsources ($n_l=2^k$, $k=1,2,3$ existing in
the source $M=100$ GeV for $(a)$ "Split" and $(b)$ "Indep" types of
sources, as discussed in text. In $(c)$ we show examples of BEC
pattern for $2$ "Split" type of sources moving apart with constant
momentum difference $\delta p = 0$, $10$ and $60$ GeV/c (achieved by
assuming in CAS first rank cascade parameters $r_1=r_2 = 0.5$,
$0.4975$ and $0.4$, respectively).} 
  \label{Fig4}
\end{figure}
results in the similar "sizes" $R$ (defined as before) with $C_2(Q=0)
- 1 = \lambda$ falling dramatically with increasing $k$ (roughly like
$1/2^k$, i.e., inversely with the number of subsources, $n_l$, as
expected from \cite{BSWW}), the former case (in which particle loose
memory of which subsource they are coming from) leads to roughly the same
$C_2(Q=0)$ but the "size" $R$ is now increasing
substantially\footnote{It is equal to, respectively, $0.87$ fm, 
$1.29$ fm, $1.99$ fm and $3.35$ fm for $P=0.5$ and $0.57$ fm, $3.26$
fm, $4.01$ and $5.59$ fm for the "minimal" weight $P_C$ given by
(\ref{eq:CASMaxEnt}).}. This is again entirely due to the fact that
in the "Split" type of source one has higher concentration of
particles in the elementary emitting cells rather then bigger number
of such cells. This results in smaller average $Q$, and this in turn
leads to bigger $R$. A special type of "Split" source is shown in
Fig. 3c. In it two initial sources (of equal masses) have from the
beginning a well defined difference in momenta, $\delta p$
(corresponding to branching parameter $r_1=r_2 = \frac{1}{2}\sqrt{1 -
(\delta p/M)^2}$ ), modelling in this way a possible influence of some
collective flow existing in the system (the total energy remains
always the same and equal to $M$, here $M=100$ GeV). Notice that,
contrary to the normal expectations, the bigger is the "flow" the
smaller is "radius" parameter $R$ obtained from the typical
exponential fit mentioned above. This is because "flow" results in
our case in smaller number of particles in the elementary emitting
cells. 

Actually dependence on the expected BEC pattern on the number and
type of subsources formed in the process of hadronization is very
important and interesting feature of our model. It allows to
understand the increase of the extracted "size" parameter $R$ with
the nuclear number $A$ in nuclear collisions. That is because with
increasing $A$ the number of collided nucleons, which somehow must
correspond to the number of sources in our case, also increases. If
they turn out to be of the "Split" type, the increase of $R$ follows
then naturally. On the contrary, for the independently treated
sources the density of particles subjected to our algorithm does not
change, hence the average $Q$ and $R$ remain essentially the same.
However, because in this case the influence of pairs of particles
from different subsources increases, the effective $\lambda =
C_2(0)-1$ now decreases substantially (as was already observed in
\cite{BSWW}). Our "Indep" type sources can therefore be used as a
possible explanation of the so called inter-$W$ BEC problem, i.e.,
the fact that essentially no BEC is being observed between pions
originating from a different $W$ in fully $W^+W^-$ final states 
\cite{WW}. This phenomenon can be understood in our model by 
assuming that produced $W$'s should be treated as "Indep" type
sources for which $\lambda$ falls dramatically.

It also allows to attempt to fit (even using such unsophisticated
hadronization model as CAS) some experimental data. As example we
\vspace{-2mm}
\begin{figure}[ht]
\noindent
  \begin{minipage}[ht]{38mm}
    \centerline{
        \epsfig{file=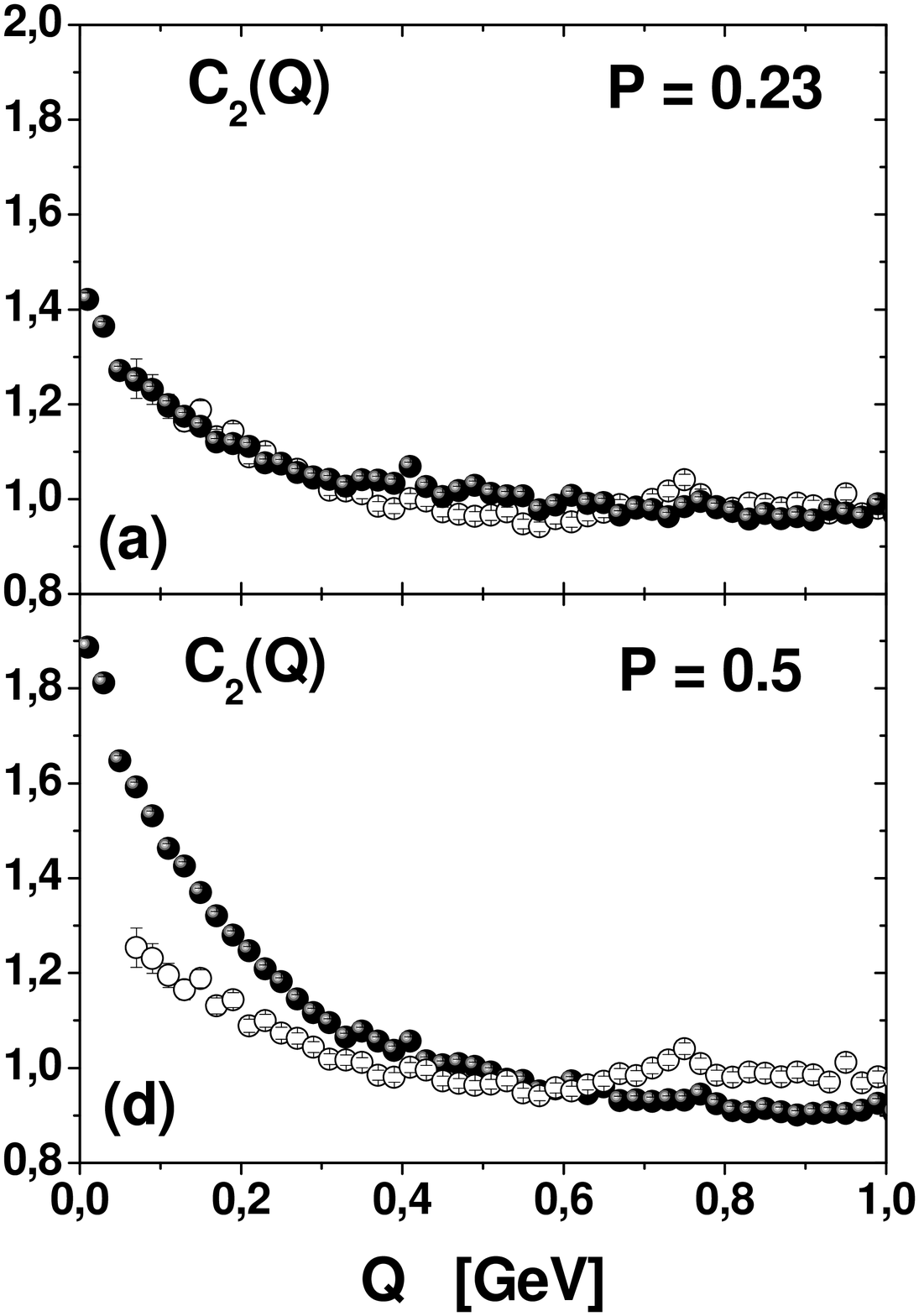, width=38mm}
     }
  \end{minipage}
\hfill
  \begin{minipage}[ht]{38mm}
    \centerline{
       \epsfig{file=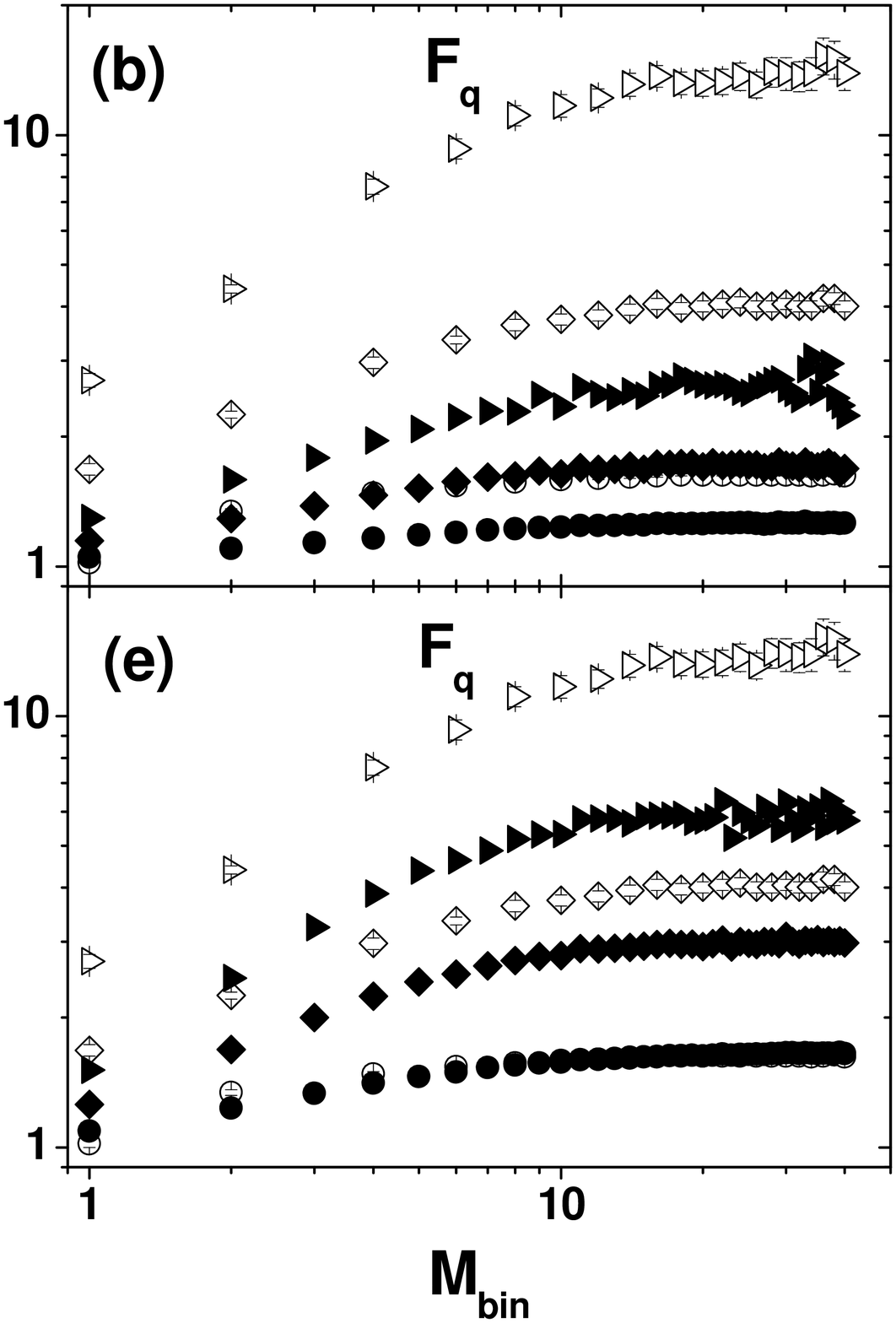, width=38mm}
     }
  \end{minipage}
\hfill
  \begin{minipage}[ht]{38mm}
    \centerline{
       \epsfig{file=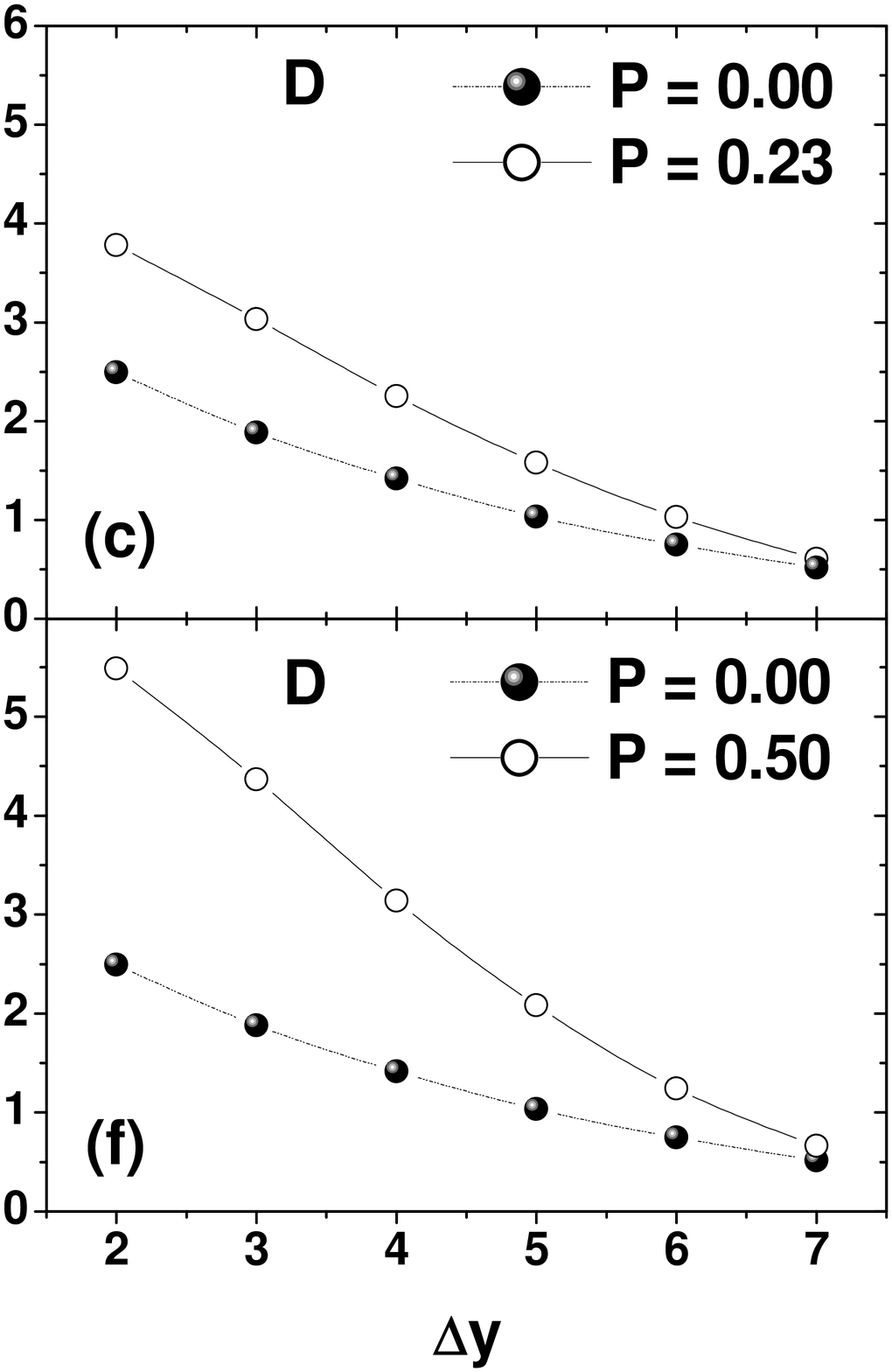, width=38mm}
     }
  \end{minipage}
\caption{Examples of results for $e^+e^-$ annihilation at $M=91.3$
GeV obtained using CAS with two subsources: $(a)$ best fit (full
symbols) to data BEC \protect{\cite{DBEC}} by DELPHI (open symbols)
obtained with $P=0.23$; $(b)$ and $(c)$ the resulting intermittency
and charge fluctuations patterns, respectively. Lower panels contain
in $(e)$ the best possible fit (full symbols) to the data on
intermittency \protect{\cite{Inter}} (open symbols, only second
moment can be reproduced, both here and in $(b)$ only $F_2$ to $F_4$
are displayed) obtained with $P=0.5$ and resulting BEC $(d)$ and
charge fluctuations patterns $(f)$ (here and in $(c)$ $P=0$
corresponds in this case to results of CAS without BEC).}   
\label{Fig5}
\end{figure}
\vspace{-2mm}
present in Fig. 5 our "best fit" to the $e^+e^-$ annihilation data on
BEC by DELPHI Collaboration \cite{DBEC} for $M=91.3$ GeV. It turns
out that such fit can be obtained only for two or more subsources
\cite{NEWBEC}. A the same figure we show also intermittency pattern
(with moments $F_q$ and $M_{bin}$ defined as in \cite{Inter})
obtained together with the BEC after application of our algorithm and 
the examples of the expected charge fluctuations in different rapidity
windows (defined as $D = 4\frac{\langle \delta Q^2\rangle}{\langle
N_{ch}\rangle}$ \cite{Qfluct}). This is done for two sets 
of parameters ($P$ and number of sources): one leading to the best
possible (which turns out very good) fit to $C_2(Q)$ (Fig. 5a) and
one leading to the best possible intermittency patter (actually only
$2$-nd moment $F_2$ can be fitted, all other moments remain still
below data indicating that intermittency connected with BEC and
provided by our algorithm as a kind of by-product, is still not the
whole effect seen in data \cite{Inter})\footnote{The charge
fluctuations $D$ are actually important for heavy ion collisions
\cite{Qfluct} and are shown here just for illustration of predictive
power of our algorithm.}.

\section{Summary and conclusions}

To summarize: we propose a new way of looking on the BEC phenomenon
observed in high energy multiparticle production processes of all
kind. Instead of cumbersome and practically very difficult (if not
outright impossible) symmetrization of the corresponding
multiparticle wave function we propose, following ideas developed in
\cite{ZAJC,OPTICS,OMT}, to look at this phenomenon as originating due
to correlations of some specific fluctuations present in such
stochastic systems as blob of hadronizing matter. As result we get
new and simple method of numerical modelling of BEC. It is based on
reassigning charges of produced particles in such a way as to make
them look like particles satisfying Bose statistics, conserves the
energy-momenta and does not alter the spatio-temporal pattern of
events or any single particle inclusive distribution (but it can
change the distributions of, separately, charged and neutral
particles leaving, however, the total distribution intact). It is
intended to generalize algorithm presented in \cite{OMT} in such a
way as to make it applicable to essentially any event generator in
which such reassignment of charges is possible. It amounts, however,
to some specific changes taking place in physical picture of the
original generator. The example of CAS is very illustrative in 
this respect. In it, at each branching vertex one has, in addition to
the energy-momentum conservation, imposed strict charge conservation
and one assumes that only $(0)\rightarrow (+-)$, $(+)\rightarrow
(+0)$ and $(-)\rightarrow (-0)$ transitions are possible. It means
that there are no multicharged vertices (i.e., vertices with multiple
charges of the same sign) in the model. However, after applying to
the finally produced particles our charge reassignment algorithm one
finds, when working the branching tree "backwards", that precisely
such vertices occur now (with charges "(++)", or "(-~-)", for
example). The total charge is, however, still conserved as are the
charges in decaying vertices (i.e., no spurious charge is being
produced because of action of our algorithm). It is plausible
therefore that to numerically get BEC pattern in an event generator
it is enough to allow in it for acumulation of charges of the same
sign at some points of hadronization procedure modelled by this
generator. This would lead, however, to extremely difficult numerical
problem with ending such algorithms without producing spurious
multicharged particles not observed in nature\footnote{It should be
noted that possibility of using multi(like)charged resonances or
clusters as possible source of BEC has been recently mentioned in
\cite{BUSH}. There remains problem of their modelling, which although
clearly visible in CAS model, as discussed here, is not so
straightforward in other approaches. However, at least in the
string-type models of hadronization, one can imagine that it could
proceed through the formation of charged (instead of neutral)
colour dipoles, i.e., by allowing formation of multi(like)charged
systems of opposite signs out of vacuum when breaking the string.
Because only a tiny fraction of such processes seems to be enough in
getting BEC in the case of CAS model, it would probably be quite
acceptable modification in the string model approach \cite{AR}. We
are indebted to late B.Andersson for very inspiring discussion at
this point at the last ISMD2001 meeting at Datong, China.}. So far
only direct pions were considered but short living resonances can
easily be included as well. The same (at least in principle) is true
in what concerns any kind of final state interactions, not mentioned
here.\\  

GW wants to mention that his interest in multiparticle production
which resulted in this presentation dates back to the seminal paper
by Professor Pokorski and L.Van Hove \cite{SP}, which spurred
formulation of the so called Interacting Gluon Model \cite{IGM}, a
simple but powerful description of high energy processes in terms of
gluonic component of hadrons. It is used (albeit in an appropriately
modified form) even at present \cite{IGM1}. He is grateful to
Professor Pokorski for his constant interest and encouragement in
this kind of research. The partial support of Polish Committee for
Scientific Research (grants 2P03B 011 18 and 621/ E-78/
SPUB/CERN/P-03/DZ4/99) is acknowledged. 


\end{document}